\documentclass[onecolumn,showpacs,preprintnumbers,floatfix,amsmath,amssymb,aps]{revtex4}
\usepackage{bm,tikz}
\usepackage{color}

\usepackage{graphicx,subfig,epsfig,epsf,color}
\usepackage[utf8x]{inputenc}
\usepackage{amssymb,amsmath,appendix}
\usepackage{slashed}
\usepackage{array}
\usepackage{hyperref}
\usepackage{ulem}

\begin{document}
\title{Constraining the mass of fermionic dark matter from its feeble interaction with hadronic matter via dark mediators in neutron stars}

\author{Atanu Guha$^{1}$ and Debashree Sen$^2$}

\affiliation{$^1$Department of Physics, Chungnam National University,\\ 99, Daehak-ro, Yuseong-gu, Daejeon-34134, South Korea}
\affiliation{$^2$Center for Extreme Nuclear Matters (CENuM), Korea University, Seoul 02841, Korea}

\email{atanu@cnu.ac.kr, debashreesen88@gmail.com}

\date{\today}




\begin{abstract}

Considering ten well-known relativistic mean field models, we invoke feeble interaction between hadronic matter and fermionic dark matter (DM) $\chi$ via new physics scalar ($\phi$) and vector ($\xi$) mediators in neutron star core, thereby forming DM admixed neutron stars (DMANSs). The chosen masses of the DM fermion ($m_{\chi}$) and the mediators ($m_{\phi}$ and $m_{\xi}$) are consistent with the self-interaction constraint from Bullet cluster while their respective couplings ($y_{\phi}$ and $y_{\xi}$) are also constrained by the present day relic abundance. Assuming that both $\phi$ and $\xi$ contribute equally to the relic abundance, we compute the equation of state of the DMANSs and consequently their structural properties. We found that for a particular (constant) DM density, the presence of lighter DM results in more massive DMANSs with larger radius. In the light of the various recent constraints like those from the massive pulsar PSR J0740+6620, the gravitational wave (GW170817) data and the results of NICER experiments for PSR J0030+0451 and PSR J0740+6620, we provide a bound on $m_{\chi}$ within the framework of the present work as $m_{\chi}\approx$ (0.1 $-$ 30) GeV for a wide range of fixed DM Fermi momenta $k_F^{\chi}$=(0.01 $-$ 0.07) GeV. In the case of the hadronic models that yield larger radii corresponding to the low mass neutron stars in the no-DM scenario, interaction with comparatively heavier DM fermion is necessary in order to ensure that the DMANSs obtained with such models satisfy the radius constraints from both GW170817 and NICER data for PSR J0030+0451.

\end{abstract}




\maketitle



\section{Introduction}
\label{Intro}

Neutron star cores are one of the most interesting, exotic and complex systems to study. At present we lack concrete experimental data at high density relevant to neutron stars and hence the knowledge regarding the composition, interaction and the equation of state of neutron star matter is based on theoretical modeling along with the related uncertainties. However, several recent astrophysical observations like those from the massive pulsar PSR J0740+6620 \cite{Fonseca:2021wxt,Miller:2021qha,Riley:2021pdl}, gravitational wave (GW170817) data \cite{LIGOScientific:2018cki} and the NICER data for PSR J0030+0451 \cite{Riley:2019yda,Miller:2019cac} put certain constraints on the equation of state of neutron stars.

 The extreme conditions of the neutron star environment is not only related to density but also gravity. The strong gravity of the neutron stars gives rise to the phenomenon of accretion and the neutron stars accrete matter from its surroundings which may include dark matter (DM) and thus forming DM admixed NSs (DMANSs). DM may also be produced in neutron star cores from neutron decay \cite{Husain:2022bxl, Husain:2022brl}. Observational evidences like the rotation curves of the galaxies, gravitational lensing, X-ray analysis of Bullet cluster \cite{Bertone:2004pz,Aghanim:2018eyx} support the existence of DM in the Universe. The Cosmic Microwave Background anisotropy maps, obtained from the Wilkinson Microwave Anisotropy Probe (WMAP) data \cite{WMAP:2012fli, Ade:2013zuv, Bennett:2012zja}, provides the present day thermal relic abundances of DM as $\sim \Omega h^2 \approx 0.12$ \cite{ParticleDataGroup:2018ovx, Bauer:2017qwy, Tanabashi:2018oca, Plehn:2017fdg, Cannoni:2015wba}. Thus the presence of DM in neutron star surroundings and eventual accretion onto neutron stars may be possible. Although the properties and the interaction of DM candidates are inconclusive, but literature suggests that the Weakly Interacting Massive Particles (WIMPs) are the most suitable DM particle candidates. Several direct detection experiments like superCDMS \cite{Agnese:2018col}, XENON100 \cite{Aprile:2012nq}, XENON1T \cite{Aprile:2018dbl}, LUX \cite{Akerib:2012ak}, PANDAX-II \cite{Wang:2020coa}, DARKSIDE-50 \cite{Agnes:2018oej}, SENSEI \cite{Crisler:2018gci} and very recently the LUX-ZEPLIN (LZ) \cite{LZ:2022ufs} etc. for the WIMPs search are being attempted worldwide. Moreover, DM may be self-interacting and in such cases the masses of the DM fermion and the DM mediators are constrained by the self-interaction constraints from bullet cluster \cite{Randall:2007ph, Bradac:2006er, Tulin:2013teo, Tulin:2017ara, Hambye:2019tjt}. Also, the self-interaction couplings are constrained to reproduce the observed non-baryonic relic density \cite{Belanger:2013oya, Gondolo:1990dk, Guha:2018mli}.

 If we consider the process of DM accretion as a source of presence of DM inside neutron star cores, then the accreted DM particles undergo collisions with the hadronic matter in the neutron stars and hence lose kinetic energy. With due course of time, the DM particles end up being gravitationally bound to the star. Eventually, the accretion stops and the DM particles attain thermal equilibrium among themselves due to the self interactions \cite{Bell:2019pyc, Bell:2020lmm}. Due to this fact, it is justified to consider the DM particle density $\rho_{\chi}$ to be almost constant \cite{Panotopoulos:2017idn,Guha:2021njn,Sen:2021wev} in the case of DMANSs. {Apart from accretion, there are other mechanisms which can be responsible for the presence of DM in neutron stars. For example, DM can be inherited by neutron stars from progenitors during supernovae explosions \cite{Nelson:2018xtr,Ciarcelluti:2010ji} or or DM may be produced inside neutron stars via neutron decay \cite{Baym:2018ljz,Husain:2022bxl,Husain:2022brl}. In the later mechanism the DM number density is reasonably high, $\rho_\chi = \left(0.01\rm{-}0.1 \right)\rho $, where, $\rho$ is total baryon number density \cite{Baym:2018ljz,Husain:2022brl}. In the present work we do not actually focus on any particular mechanism as the possible source of the presence of DM in neutron stars. We intend to show how the presence of DM, through any of the possible mechanisms as discussed above, can affect the structural properties of neutron stars. Also in this context we investigate the possible range of DM mass in order to satisfy the observational constraints on neutron star properties}. Co-existing along with the baryonic matter inside neutron star cores, DM may or may not interact with hadronic matter. In case they do not interact, the two types of matter coexist in the two fluid form \cite{Lopes:2018oao, Ellis:2018bkr, Li:2012ii, Tolos:2015qra, Deliyergiyev:2019vti, Rezaei:2016zje, Mukhopadhyay:2016dsg, Mukhopadhyay:2015xhs, Panotopoulos:2017pgv, Jimenez:2021nmr, Panotopoulos:2018ipq, Leung:2022wcf, Karkevandi:2021ygv, Lourenco:2021dvh, Gleason:2022eeg, Dengler:2021qcq, Panotopoulos:2018joc, Panotopoulos:2017eig, Miao:2022rqj, Rutherford:2022xeb} whereas the interaction between the DM and the baryonic matter is also suggested by \cite{Panotopoulos:2017idn, Bertoni:2013bsa, Nelson:2018xtr, Bhat:2019tnz, Lourenco:2022fmf, Dutra:2022mxl, Hong:2023udv, Mu2023}, mostly via the Higgs boson as mediator. The interaction between DM and the hadronic matter of the star must be extremely weak \cite{Zheng:2016ygg} to prevent the collapse of the star into a black hole due to heavy accretion of DM. Therefore in \cite{Sen:2021wev} we invoked feeble interaction between hadronic and fermionic DM $\chi$ via a new scalar mediator $\phi$ and a dark vector mediator $\xi$ in \cite{Guha:2021njn} in order to explain the possible existence of DMANSs. $\phi$ and $\xi$ interact with the hadronic matter $\psi$ with a very feeble coupling strength. The masses of DM fermion $m_{\chi}$ and the mediators ($m_{\phi}$ and $m_{\xi}$) and the couplings ($y_{\phi}$ and $y_{\xi}$) are consistent with the self-interaction constraint from the Bullet cluster and from the present day relic abundance, respectively. In both \cite{Sen:2021wev, Guha:2021njn} we considered only the effective chiral model as the hadronic model to study the effects of DM interaction on the properties of DMANSs. We concluded that mass of DM ($m_{\chi}$) plays a very important role in determining the structural properties of DMANSs. The massive the DM, the less are the maximum mass, radius and tidal deformability of the DMANSs. In the present work we aim to constrain the value of $m_{\chi}$ or rather we seek a possible range of $m_{\chi}$ for which the DMANSs satisfy the constraints on the structural properties of compact stars obtained from PSR J0740+6620, GW170817 and the NICER data for PSR J0030+0451. For the purpose we consider ten well-known relativistic mean field (RMF) models viz. TM1, GM1, NL3, PK1, DD-MEX, DD2, TW99, DD-ME2, PK-DD and DD-LZ1. In order to obtain the range of $m_{\chi}$, we consider one minimum and another maximum value of the DM Fermi momentum $k_F^{\chi}$ which gives the maximum and minimum values of the constant DM particle density $\rho_{\chi}$.

 This paper is organized as follows. In the next section \ref{Formalism}, we briefly address the framework of the ten RMF hadronic models. In the same section, we also discuss the mechanism of invoking the DM interaction with hadronic matter via the dark mediators $\phi$ and $\xi$ and the the structural properties of the DMANSs. We then present our results and corresponding discussions in section \ref{Results}. We summarize and conclude in the final section \ref{Conclusion} of the paper.
  

\section{Formalism}
\label{Formalism}

 Following our previous works \cite{Sen:2021wev, Guha:2021njn} we introduce feeble interaction of the dark fermion ($\chi$) with the hadronic matter ($\psi$=n, p) through the scalar ($\phi$) and vector ($\xi$) new physics mediators in neutron star core. For the pure hadronic matter sector we consider ten well-known RMF models of two different classes - i) models with non-linear self couplings like TM1 \cite{Sugahara:1993wz}, GM1 \cite{Glendenning:1991es}, NL3 \cite{Lalazissis:1996rd}, and PK1 \cite{Long:2003dn} and ii) models with density-dependent couplings like DD-MEX \cite{Taninah:2019cku}, DD2 \cite{Typel:2009sy}, DD-ME2 \cite{Lalazissis:2005de}, PK-DD \cite{Long:2003dn}, DD-LZ1 \cite{Wei:2020kfb}, and TW99 \cite{Lu:2011wy}. For the dark sector, we consider the phenomenological treatment to describe the self-interaction of non-relativistic DM by a Yukawa potential \cite{Tulin:2013teo}

\begin{eqnarray}
V(r) = \pm \frac{\alpha_\chi}{r} e^{-m_\phi r}
\end{eqnarray} 

where, $\alpha_\chi = \frac{y^2}{4 \pi}$ is the dark fine structure constant. We consider that $\phi$ and $\xi$ have their respective couplings as $y_{\phi}$ and $y_{\xi}$ with $\chi$ 

\begin{eqnarray} 
\mathcal{L}_{int} = \begin{cases}
y_\phi \phi \bar{\chi} \chi \\
y_\xi \bar{\chi} \gamma_\mu \chi \xi^\mu
\end{cases} 
\end{eqnarray} 

The complete Lagrangian is given as

\begin{eqnarray} 
\mathcal{L}=\bar{\psi}[\gamma_{\mu}(i\partial^{\mu} -g_{\omega}\omega^{\mu} -g_\rho \vec{\rho_\mu}\cdot \vec{\tau} -g_{\xi}\xi^{\mu}) -(M +g_{\sigma}\sigma +g_{\phi}\phi)]\psi
+ \frac{1}{2}\partial_{\mu}\sigma\partial^{\mu} -\frac{1}{2}m_{\sigma}^2\sigma^2 -\frac{1}{3}g_2\sigma^3 -\frac{c}{4}g_3\sigma^4  \nonumber\\
-\frac{1}{4}\omega_{\mu\nu}\omega^{\mu\nu} +\frac{1}{2}m_{\omega}^2\omega_{\mu}\omega^{\mu} +\frac{1}{4}c_3(\omega_{\mu}\omega^{\mu})^2
-\frac{1}{4} \vec{R}_{\mu \nu} \cdot \vec{R}^{\mu \nu} +\frac{1}{2} m_\rho^2 \vec{\rho_\mu}\cdot \vec{\rho^\mu} \nonumber\\
+\frac{1}{2} \partial_\mu \phi \partial^\mu \phi - \frac{1}{2}m_\phi^2 \phi^2 -\frac{1}{4} V_{\mu \nu} V^{\mu \nu} +\frac{1}{2} m_\xi^2 \xi_\mu \xi^\mu
+\bar{\chi} \left[\left( i \gamma_\mu \partial^\mu - y_\xi \gamma_\mu \xi^\mu \right) -\left(m_\chi + y_\phi \phi \right) \right] \chi 
\label{eq:rmf_lagrangian} 
\end{eqnarray} 

In the pure hadronic sector the nucleons interact via the scalar $\sigma$, the vector $\omega$ and iso-vector $\rho$ mesons. The vacuum expectation values (VEVs) of the meson fields ($\sigma_0$, $\omega_0$ and $\rho_{03}$) in RMF approximation remain unaffected due to the presence of DM and the expressions can be found in \cite{Xia:2022dvw}. The mesons in the hadronic sector have density independent couplings $g_{\sigma}$, $g_{\omega}$ and $g_{\rho}$ with the nucleons for the models like TM1, GM1, NL3, and PK1. For such models $g_2$ and $g_3$ are the higher order scalar field coefficients while $c_3$ is the higher order vector field coefficient. These non-linear self couplings are effectively considered in order to account for the in-medium effects. In the following table we first show the density independent couplings ($g_{\sigma}$, $g_{\omega}$, $g_{\rho}$, $g_2$, $g_3$, and $c_3$) and the mass of mesons ($m_{\sigma}$, $m_{\omega}$ and $m_{\rho}$) and neutron ($m_n$) and proton ($m_p$) adopted in the models like TM1 \cite{Sugahara:1993wz}, GM1 \cite{Glendenning:1991es}, NL3 \cite{Lalazissis:1996rd}, and PK1 \cite{Long:2003dn} according to the respective references.

\begin{table}[!ht]
\caption{The density independent meson-nucleon couplings and parameters adopted in the models TM1 \cite{Sugahara:1993wz}, GM1 \cite{Glendenning:1991es}, NL3 \cite{Lalazissis:1996rd}, and PK1 \cite{Long:2003dn}.}
\setlength{\tabcolsep}{5.0pt}
\begin{tabular}{cccccccccccc}
\hline
\hline
Model & $m_n$ & $m_p$ & $m_{\sigma}$ & $m_{\omega}$ & $m_{\rho}$ & $g_{\sigma}$ & $g_{\omega}$ & $g_{\rho}$ & $g_2$ & $g_3$ & $c_3$  \\
& (MeV) & (MeV) & (MeV) & (MeV) & (MeV) & & & & $({\rm fm}^{-1})$ &  \\
\hline
TM1 & 938 & 938 & 511.198 & 783 & 770 & 10.0289 & 12.6139 & 4.6322 & -7.2325 & 0.6183 & 71.3075 \\
GM1 & 938 & 938 & 510 & 783 & 770 & 8.874 43 & 10.60957 & 4.09772 & -9.7908 & -6.63661 & 0 \\
NL3 & 939 & 939 & 508.1941 & 782.501 & 763 & 10.2169 & 12.8675 & 4.4744 & -10.4307 & -28.8851 & 0 \\
PK1 & 939.5731 & 938.2796 & 514.0891 & 784.254 & 763 & 10.3222 & 13.0131 & 4.5297 & -8.1688 & -9.9976 & 55.636 \\
\hline
\hline
\end{tabular}
\label{tab:1}
\end{table}

In models like DD-MEX \cite{Taninah:2019cku}, DD2 \cite{Typel:2009sy}, DD-ME2 \cite{Lalazissis:2005de}, PK-DD \cite{Long:2003dn}, DD-LZ1 \cite{Wei:2020kfb}, and TW99 \cite{Lu:2011wy} $g_2$=$g_3$=$c_3$=0 and the in-medium effects are treated with the density-dependent couplings following the Typel-Wolter ansatz \cite{Lu:2011wy} as

\begin{eqnarray}
g_i(\rho)=g_i a_i \frac{1+b_i(x+d_i)^2}{1+c_i(x+d_i)^2}
\end{eqnarray}

where $i=\sigma, \omega$ and $x=\rho/\rho_0$ while

\begin{eqnarray}
g_{\rho}(\rho)=g_{\rho} \rm{exp} [a_{\rho}(x-1)]
\end{eqnarray}

\begin{table}[!ht]
\caption{The density dependent meson-nucleon couplings and parameters adopted in the models DD-MEX \cite{Taninah:2019cku}, DD2 \cite{Typel:2009sy}, DD-ME2 \cite{Lalazissis:2005de}, PK-DD \cite{Long:2003dn}, DD-LZ1 \cite{Wei:2020kfb}, and TW99 \cite{Lu:2011wy}.}
\setlength{\tabcolsep}{5.0pt}
\begin{tabular}{cccccccccccccccccc}
\hline
\hline
Model & $m_n$ & $m_p$ & $m_{\sigma}$ & $m_{\omega}$ & $m_{\rho}$ & $g_{\sigma}$ & $g_{\omega}$ & $g_{\rho}$ \\
& (MeV) & (MeV) & (MeV) & (MeV) & (MeV) & & &\\
\hline
DD-MEX & 938.5 & 938.5 & 547.3327 & 783 & 763 & 10.706722 & 13.338846 & 3.619020 \\
DD2 & 939.56536 & 938.27203 & 546.212459 & 783 & 763 & 10.686681 & 13.342362 & 3.626940 \\
DD-ME2 & 938.5 & 938.5 & 550.1238 & 783 & 763 & 10.5396 & 13.0189 & 3.6836 \\ 
PK-DD & 939.5731 & 938.2796 & 555.5112 & 783 & 763 & 10.7385 & 13.1476 & 4.2998 \\
DD-LZ1 & 938.9 & 938.9 & 538.619216 & 783 & 763 & 12.001429 & 14.292525 & 7.575467 \\
TW99 & 939 & 939 & 550 & 783 & 763 & 10.7285 & 13.2902 & 3.6610 \\
\hline
Model & $a_{\sigma}$ & $b_{\sigma}$ & $c_{\sigma}$ & $d_{\sigma}$ & $a_{\omega}$ & $b_{\omega}$ & $c_{\omega}$ & $d_{\omega}$ & $a_{\rho}$ \\
\hline
DD-MEX & 1.397043 & 1.334964 & 2.067122 & 0.401565 & 1.393601 & 1.019082 & 1.605966 & 0.455586 & 0.620220 \\
DD2 & 1.357630 & 0.634442 & 1.005358 & 0.575810 & 1.369718 & 0.496475 & 0.817753 & 0.638452 & 0.983955 \\
DD-ME2 & 1.3881 & 1.0943 & 1.7057 & 0.4421 & 1.3892 & 0.9240 & 1.4620 & 0.4775 & 0.5647 \\
PK-DD & 1.327423 & 0.435126 & 0.691666 & 0.694210 & 1.342170 & 0.371167 & 0.611397 & 0.738376 & 0.183305 \\
DD-LZ1 & 1.062748 & 1.763627 & 2.308928 & 0.379957 & 1.059181 & 0.418273 & 0.538663 & 0.786649 & 0.776095 \\
TW99 & 1.365469 & 0.226061 & 0.409704 & 0.901995 & 1.402488 & 0.172577 & 0.344293 & 0.983955 & 0.515000 \\
\hline
\hline
\end{tabular}
\label{tab:2}
\end{table}

All the relevant masses and the parameters are listed in Tables~\ref{tab:1} and \ref{tab:2}. The saturation properties like the saturation density $\rho_0$, symmetry energy $J_0$, slope $L_0$, nuclear incompressibility $K_0$, skewness coefficient $S_0$, and the curvature parameter $K_{sym}$ of the nuclear symmetry energy as obtained for all the above ten models considered in this present work for the specific parameters can be found in the respective references and also in \cite{Xia:2022dvw}.

The dark bosons $\phi$ and $\xi$ interact with the hadronic matter $\psi$ with a very feeble coupling strength $g_{\phi}=g_{\xi}\sim$10$^{-4}$. The VEVs of the DM mediator fields in RMF approximation are

\begin{eqnarray} 
\phi_0=\frac{m^\star_\chi-m_\chi}{y_{\phi}}
\end{eqnarray} 

and

\begin{eqnarray} 
\xi_0 = \frac{g_\xi \rho + y_\xi \rho_\chi}{m_\xi^2}
\end{eqnarray}

The modified effective mass due to DM interaction is 

\begin{eqnarray} 
m_B^{\star}=M_B+g_{\sigma}\sigma +g_{\phi}\phi
\end{eqnarray} 

while the modified chemical potential is

\begin{eqnarray} 
\mu_B=\sqrt{k_B^2 + {m_B^{\star}}^2} + g_{\omega}\omega_0 + g_{\rho}I_{3B}\rho_{03} + \Sigma^R + g_{\xi}\xi_0
\end{eqnarray} 

where, the rearrangement term $\Sigma^R$=0 for the models with density independent couplings and for the models with density dependent couplings it is given by \cite{Lenske:1995wyj} as

\begin{eqnarray} 
\Sigma^R = \frac{dg_{\sigma}(\rho)}{d\rho}\sigma_0\rho_{SB} + \frac{dg_{\omega}(\rho)}{d\rho}\omega_0\rho + \frac{dg_{\rho}(\rho)}{d\rho}\rho_{03}I_{3B}\rho_B
\end{eqnarray} 

Here, $B$=n, p and $\rho_S$ is the scalar density. $I_{3B}$ is the third component of isospin for the individual nucleons.

The complete expressions for the equation of state is also modified due to the presence of DM. The energy density $\varepsilon$ is given as

\begin{eqnarray}
\varepsilon=\frac{1}{2}m_{\sigma}^2\sigma_0^2 +\frac{1}{3}g_2\sigma_0^3 +\frac{1}{4}g_3\sigma_0^4 +\frac{1}{2}m_{\omega}^2\omega_0^2 +\frac{3}{4}c_3\omega_0^4 +\frac{1}{2}m_{\rho}^2\rho_{03}^2 \nonumber \\ +\frac{\gamma}{2\pi^2} \sum_{B=n,p} \int_0^{k_F} \sqrt{k_B^2 + {m_B^{\star}}^2}~k_B^2 dk + \frac{\gamma}{2 \pi^2} \sum_{l=e,\mu} \int_0^{k_l} \sqrt{k_l^2 + m_l^2}~ k_l^2 dk_l \nonumber \\ +\frac{1}{2}m_{\phi}^2\phi_0^2 +\frac{1}{2}m_{\xi}^2\xi_0^2 +\frac{\gamma_{\chi}}{2\pi^2} \int_0^{k_F^{\chi}} \sqrt{k_{\chi}^2 + {m_{\chi}^{\star}}^2}~ k_{\chi}^2 dk{\chi}
\label{e}
\end{eqnarray} 

and the pressure is given as

\begin{eqnarray}
P=-\frac{1}{2} m_{\sigma}^2\sigma_0^2 -\frac{1}{3}g_2\sigma_0^3 -\frac{1}{4}g_3\sigma_0^4 +\frac{1}{2}m_{\omega}^2\omega_0^2 +\frac{1}{4}c_3\omega_0^4 +\frac{1}{2}m_{\rho}^2\rho_{03}^2 \nonumber \\ +\frac{\gamma}{6\pi^2} \sum_{B=n,p} \int_0^{k_F} \frac{k_B^4 dk}{\sqrt{k_B^2 + {m_B^{\star}}^2}} +\frac{\gamma}{6 \pi^2} \sum_{l=e,\mu} \int_0^{k_l} \frac{k_l^4 dk_l}{\sqrt{k_l^2 + m_l^2}} \nonumber \\ +\frac{1}{2}m_{\phi}^2\phi_0^2 +\frac{1}{2}m_{\xi}^2\xi_0^2 +\frac{\gamma_\chi}{6 \pi^2} \int_0^{k_F^{\chi}} \frac{k_\chi^4 dk_\chi}{\sqrt{k_\chi^2 + {m_\chi^{\star}}^2}}
\label{p}
\end{eqnarray} 

 As mentioned in the Introduction section \ref{Intro} that following our previous works \cite{Sen:2021wev, Guha:2021njn, Sen:2022pfr}, in the present work the values of $m_{\chi}$, $m_{\phi}$ and $m_{\xi}$ are considered consistent with the self-interaction constraints from bullet cluster \cite{Randall:2007ph, Bradac:2006er, Tulin:2013teo, Tulin:2017ara, Hambye:2019tjt} while the self-interaction couplings are also chosen by reproducing the observed non-baryonic relic density \cite{Belanger:2013oya, Gondolo:1990dk, Guha:2018mli}. The permitted values of $m_{\phi}$ and $m_{\xi}$ corresponding to the range of $m_{\chi}$ are already shown in our previous works \cite{Guha:2021njn, Sen:2021wev, Sen:2022pfr}.

 With the obtained DMANS equation of state, we compute the structural properties like the gravitational mass ($M$) and the radius ($R$) of the DMANSs in static conditions  by integrating the Tolman-Oppenheimer-Volkoff (TOV) equations \cite{Tolman:1939jz,Oppenheimer:1939ne}. The dimensionless tidal deformability ($\Lambda$) is obtained in terms of the mass, radius and the tidal love number ($k_2$) following \cite{Hinderer:2007mb,Hinderer:2009ca}.
 
 {In the present work we have considered the DM number density $\rho_{\chi}$ to be constant via constant DM Fermi momentum throughout the radial profile of the star following \cite{Panotopoulos:2017idn} and our previous works \cite{Guha:2021njn,Sen:2021wev}. This number density is quite high compared to the density considered in \cite{Bell:2020obw} where the authors have shown that using the quark-meson coupling (QMC) model and considering local DM mass density (=0.3 GeV/cc), the capture rate of accreted DM can be $\sim$(10$^{33}$ - 10$^{43}$) GeV s$^{-1}$ for DM mass $m_{\chi}\sim$ 1 GeV with different operators and for different neutron star mass. Further from \cite{Bell:2020obw} we find that the DM capture rate is roughly directly proportional to the DM number density $\rho_{\chi}$. In the present work we have the DM number density as 4.4$\times$10$^{-6}$ fm$^{-3}$ (mass density 4.4$\times$10$^{33}$ GeV/cc) and 9.5$\times$10$^{-4}$ fm$^{-3}$ (mass density 9.5$\times$10$^{35}$ GeV/cc) for Fermi momentum $k_F^{\chi}$ to be 0.01 GeV and 0.07 GeV, respectively. So for $m_{\chi}\sim$ 1 GeV the DM mass density is very high compared to the local DM density. Therefore for the case where the accretion is the only mechanism for presence of DM in neutron stars, the DM capture rate has to be enhanced compared to \cite{Bell:2020obw} roughly by an factor of $\sim$10$^{34}$ for $k_F^{\chi}$=0.01 GeV and $\sim$10$^{37}$ for $k_F^{\chi}$=0.07 GeV, to explain such a high density of DM inside neutron star. So in the present work the maximum DM capture rate is $\sim$10$^{77}$ GeV s$^{-1}$ for $k_F^{\chi}$=0.01 GeV and $\sim$10$^{80}$ GeV s$^{-1}$ for $k_F^{\chi}$=0.07 GeV. Considering our estimate of DM capture rate, we find that it is largely inconsistent with the results of \cite{Bell:2020obw}. The main reason is that our consideration of $\rho_\chi$ is quite high which leads to high values of the DM capture rate. A possible solution to fix this problem  may be to consider the local DM density. It can be expected that consideration of the local DM density can match the order of DM capture rate as obtained by \cite{Bell:2020obw}.}
 
 { However, as mentioned in the introduction, there maybe other possible sources for the presence of DM inside neutron stars. So even if the DM density in the vicinity of the neutron star is considered to be the local DM density, which can make the capture rate to be consistent with \cite{Bell:2020obw}, the DM density inside the neutron star can be quite high due to the other mechanisms involved, as seen from \cite{Baym:2018ljz,Husain:2022brl}. This maybe another feasible explanation for the high DM density inside neutron star along with the DM capture rate being consistent with \cite{Bell:2020obw}.}
 
 { So irrespective of the mechanism of the presence of DM inside neutron star, we proceed to study the effects of DM on the structural properties of neutron stars in the next section.}


\section{Results}
\label{Results}

\subsection{Neutron stars without dark matter}

\begin{figure}[!ht]
\centering
\subfloat[]{\includegraphics[width=0.5\textwidth]{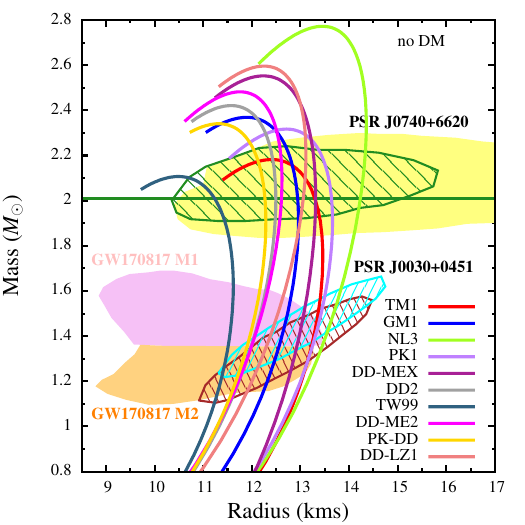}\protect\label{mr_noDM}}
\subfloat[]{\includegraphics[width=0.5\textwidth]{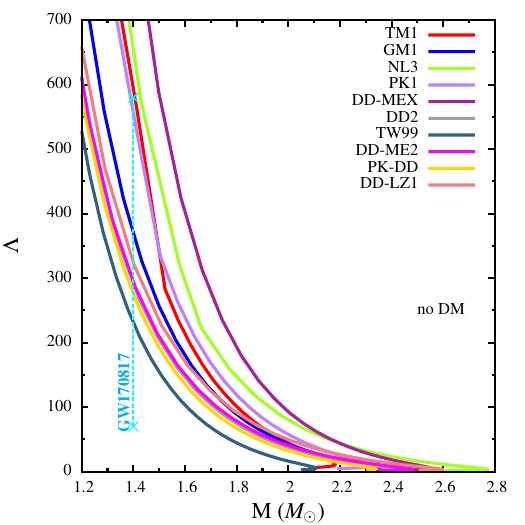}\protect\label{LamM_noDM}}
\caption{\it Variation of (a) mass with radius and (b) tidal deformability with mass of neutron stars without dark matter with hadronic models. In figure (a) the observational limits imposed from the most massive pulsar PSR J0740+6620 ($M = 2.08 \pm 0.07 M_{\odot}$) \cite{Fonseca:2021wxt} and $R = 13.7^{+2.6}_{-1.5}$ km (yellow shaded region \cite{Miller:2021qha}) or $R = 12.39^{+1.30}_{-0.98}$ km (shaded region with green diagonal lines \cite{Riley:2021pdl}) are also indicated. The constraints on $M-R$ plane prescribed from GW170817 (pink and orange shaded regions \cite{LIGOScientific:2018cki}) and the NICER experiment for PSR J0030+0451 (shaded regions with brown \cite{Riley:2019yda} and cyan \cite{Miller:2019cac} diagonal lines) are also compared. In figure (b) the constraint on $\Lambda_{1.4}$ from GW170817 \cite{LIGOScientific:2018cki} is also shown.}
\label{mLamM_noDM}
\end{figure}

We first show the results of the structural properties of neutron stars obtained with the ten chosen RMF models in the absence of DM in Fig.~\ref{mLamM_noDM}. Fig.~\ref{mr_noDM} shows the variation of mass with radius and Fig.~\ref{LamM_noDM} shows the relation of tidal deformability with mass of neutron stars without DM. It can be seen from Fig.~\ref{mr_noDM} that among the ten chosen RMF models, NL3 yields the most massive neutron star with maximum radius while with the TW99 model we obtain the least massive neutron star configuration with minimum radius compared to that obtained with the other models. The neutron star configurations obtained with all the chosen models satisfy the constraint on the mass-radius relationship of the neutron stars obtained from the most massive pulsar PSR J0740+6620 \cite{Fonseca:2021wxt, Riley:2021pdl, Miller:2021qha} and also the NICER data for PSR J0030+0451 \cite{Riley:2019yda, Miller:2019cac}. However, it is well known that the constraints from GW170817 \cite{LIGOScientific:2018cki} both on the $M-R$ and $M-\Lambda$ planes are not or barely satisfied by the results with the NL3, TM1, and the PK1 models. Our results in Figs.~\ref{mr_noDM} and \ref{LamM_noDM} support the same. The result of the DD-MEX model satisfies the bound from GW170817 in the $M-R$ plane but not in the $M-\Lambda$ plane. This subsection serves as an overview of the present literature. We present the existing results of the structural properties of neutron star without the presence of DM with different models particularly for the purpose of comparison.

\subsection{Dark matter admixed neutron stars with maximum $k_F^{\chi}$=0.07 GeV}

We next present our results of the structural properties of the DMANSs obtained with the ten chosen hadronic models, first considering maximum value of $k_F^{\chi}$=0.07 GeV. This maximum value of $k_F^{\chi}$ implies the maximum DM density $\rho_{\chi}^{max}$(= 9.5 $\times$ 10$^{-4}$ fm$^{-3}$) i.e., when DM populates the neutron star the maximum. The dark matter accreted by neutron stars affects the equation of state and consequently the structural properties of the dark matter admixed neutron stars. From equations \ref{e} and \ref{p} (last terms) it can be seen that the equation of state i.e. the total energy density and pressure of the dark matter admixed neutron star is not only affected by the dark matter Fermi momentum but also the mass of the dark matter. Since the structural properties of the star like the mass, radius and tidal deformability are directly dictated by the equation of state, the presence of dark matter and its Fermi momentum and the mass play important role in determining the structural properties of the star. In our previous works \cite{Sen:2021wev, Guha:2021njn} we found that lighter fermionic DM results in more massive DMANSs with larger radius. Therefore, we check with each model the suitable mass range of fermionic DM in order to obtain reasonable DMANSs configurations in the light of the different astrophysical constraints on the structural properties of compact stars.

\begin{figure}[!ht]
\centering
\subfloat[TM1]{\includegraphics[width=0.49\textwidth]{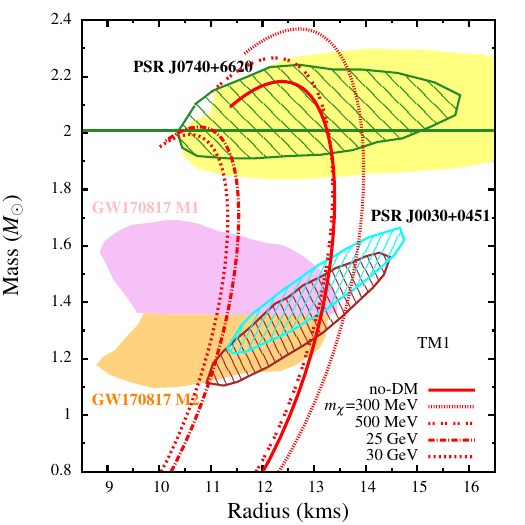}\label{mr_mchiTM1}}
\hfill
\subfloat[GM1]{\includegraphics[width=0.49\textwidth]{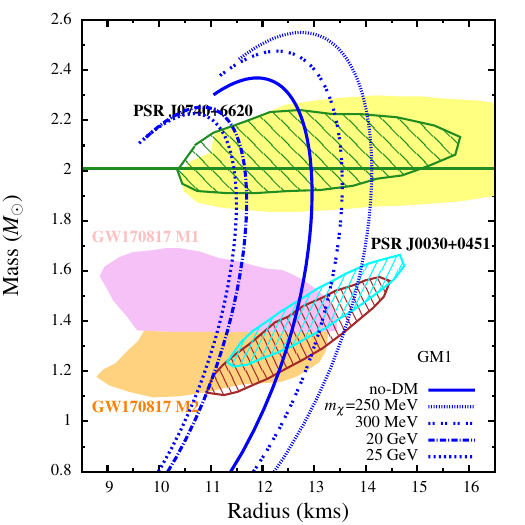}\label{mr_mchiGM1}}
\\
\centering
\subfloat[NL3]{\includegraphics[width=0.49\textwidth]{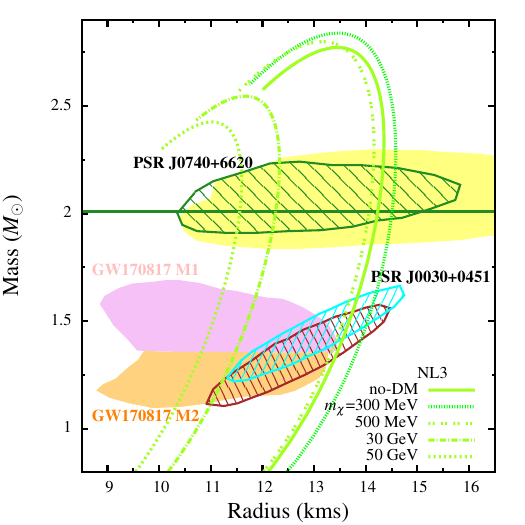}\label{mr_mchiNL3}}
\hfill
\subfloat[PK1]{\includegraphics[width=0.49\textwidth]{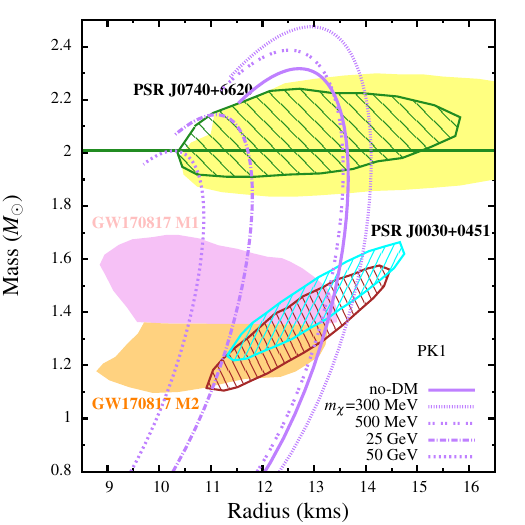}\label{mr_mchiPK1}}
\caption{\it Variation of mass with radius of dark matter admixed neutron stars for different values of $m_{\chi}$ and maximum $k_F^{\chi}$ with hadronic models (a) TM1 (b) GM1 and (c) NL3 and (d) PK1.}
\end{figure}

\begin{figure}[!ht]
\centering
\subfloat[TM1]{\includegraphics[width=0.49\textwidth]{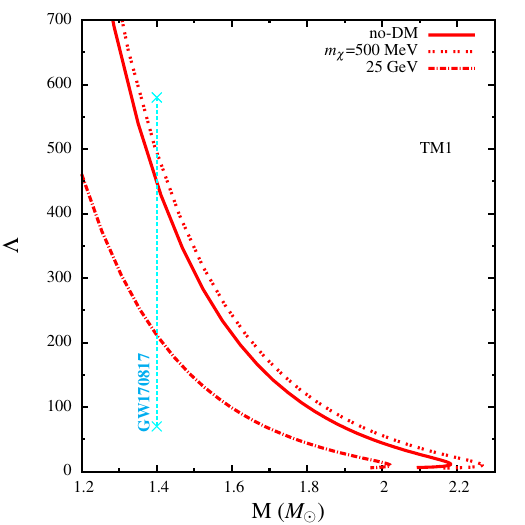}\label{LamM_mchiTM1}}
\hfill
\subfloat[GM1]{\includegraphics[width=0.49\textwidth]{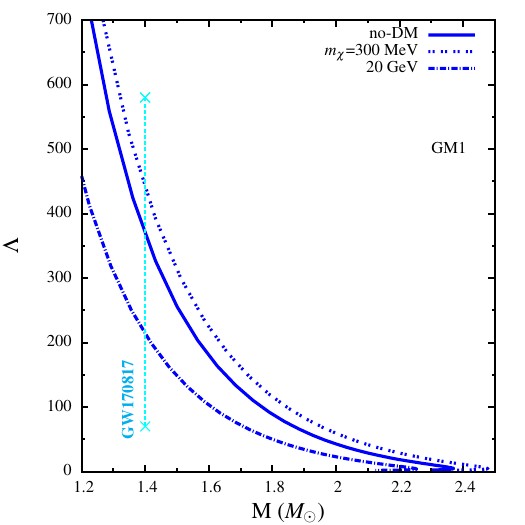}\label{LamM_mchiGM1}}
\\
\centering
\subfloat[NL3]{\includegraphics[width=0.49\textwidth]{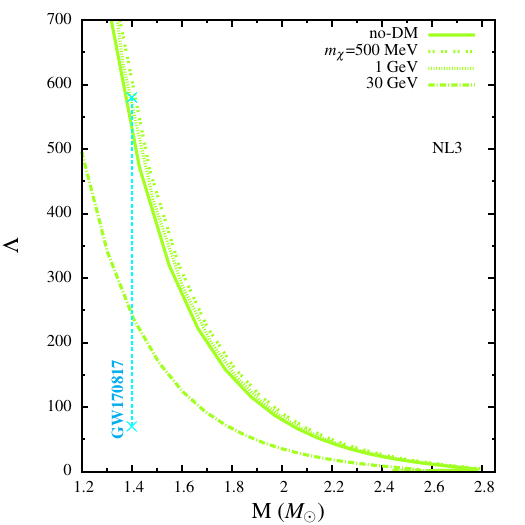}\label{LamM_mchiNL3}}
\hfill
\subfloat[PK1]{\includegraphics[width=0.49\textwidth]{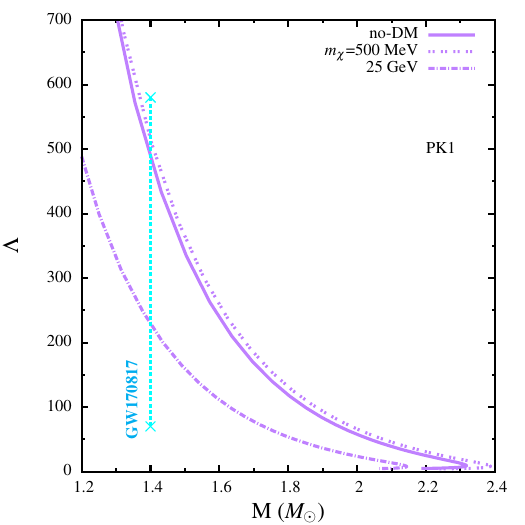}\label{LamM_mchiPK1}}
\caption{\it Variation of tidal deformability with mass of dark matter admixed neutron stars for different values of $m_{\chi}$ and maximum $k_F^{\chi}$ with hadronic models (a) TM1 (b) GM1 and (c) NL3 and (d) PK1.}
\end{figure}

\begin{figure}[!ht]
\centering
\subfloat[DD-MEX]{\includegraphics[width=0.49\textwidth]{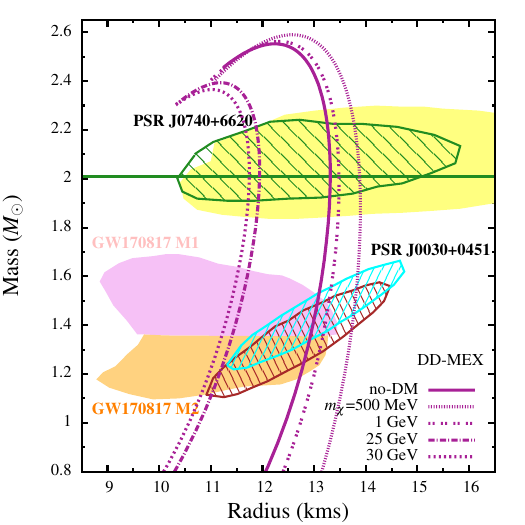}\label{mr_mchiDDMEX}}
\hfill
\subfloat[DD2]{\includegraphics[width=0.49\textwidth]{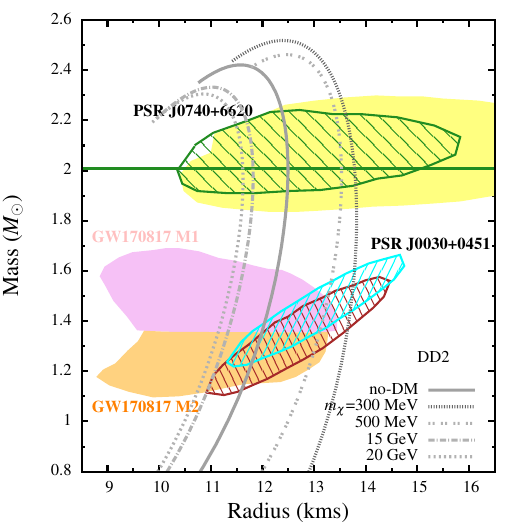}\label{mr_mchiDD2}}
\\
\centering
\subfloat[TW99]{\includegraphics[width=0.49\textwidth]{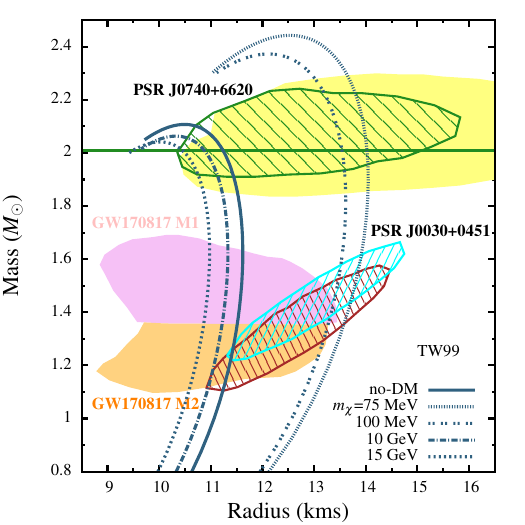}\label{mr_mchiTW99}}
\hfill
\subfloat[DD-ME2]{\includegraphics[width=0.49\textwidth]{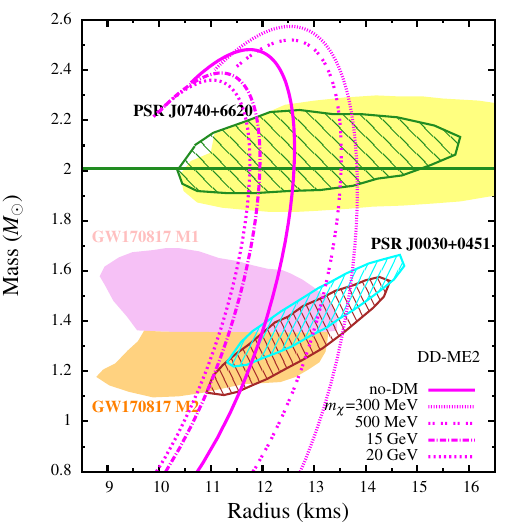}\label{mr_mchiDDME2}}
\caption{\it Variation of mass with radius of dark matter admixed neutron stars for different values of $m_{\chi}$ and maximum $k_F^{\chi}$ with hadronic models (a) DD-MEX (b) DD2 and (c) TW99 and (d) DD-ME2.}
\end{figure}

\begin{figure}[!ht]
\centering
\subfloat[DD-MEX]{\includegraphics[width=0.49\textwidth]{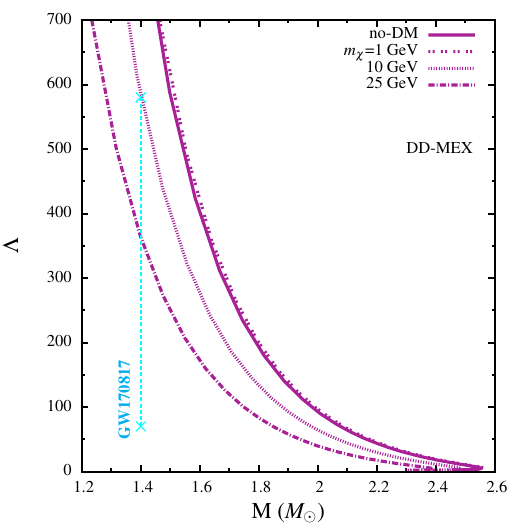}\label{LamM_mchiDDMEX}}
\hfill
\subfloat[DD2]{\includegraphics[width=0.49\textwidth]{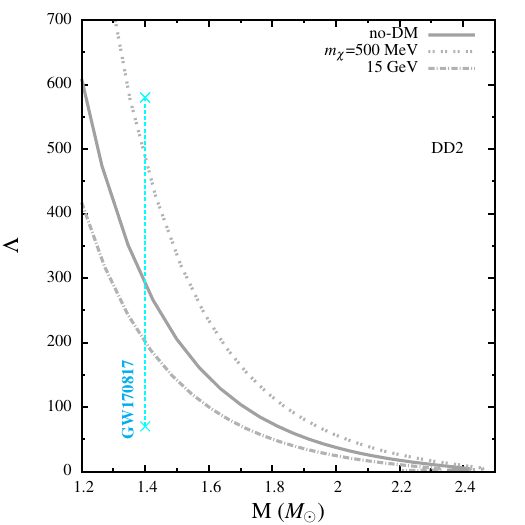}\label{LamM_mchiDD2}}
\\
\centering
\subfloat[TW99]{\includegraphics[width=0.49\textwidth]{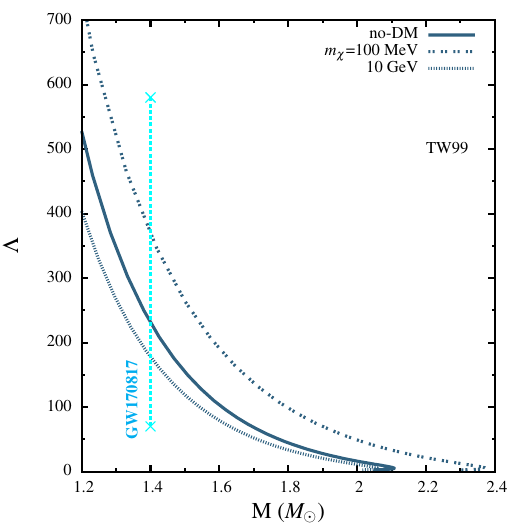}\label{LamM_mchiTW99}}
\hfill
\subfloat[DD-ME2]{\includegraphics[width=0.49\textwidth]{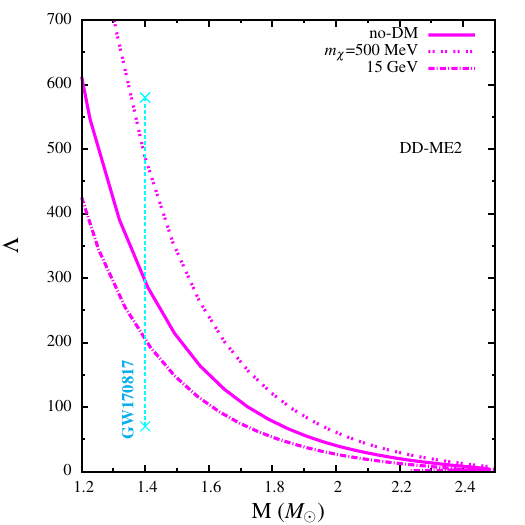}\label{LamM_mchiDDME2}}
\caption{\it Variation of tidal deformability with mass of dark matter admixed neutron stars for different values of $m_{\chi}$ and maximum $k_F^{\chi}$ with hadronic models (a) DD-MEX (b) DD2 and (c) TW99 and (d) DD-ME2.}
\end{figure}

\begin{figure}[!ht]
\centering
\subfloat[PK-DD]{\includegraphics[width=0.5\textwidth]{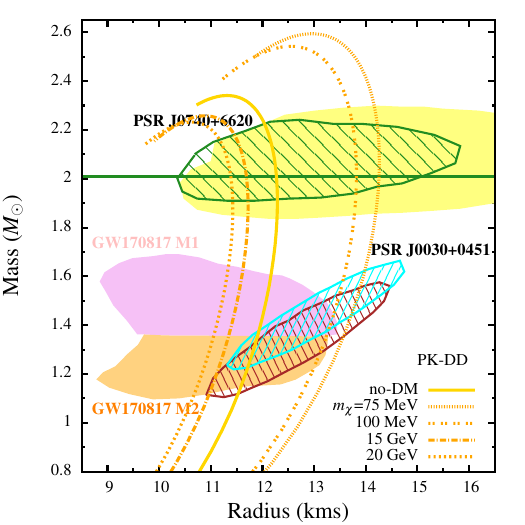}\protect\label{mr_mchiPKDD}}
\subfloat[DD-LZ1]{\includegraphics[width=0.5\textwidth]{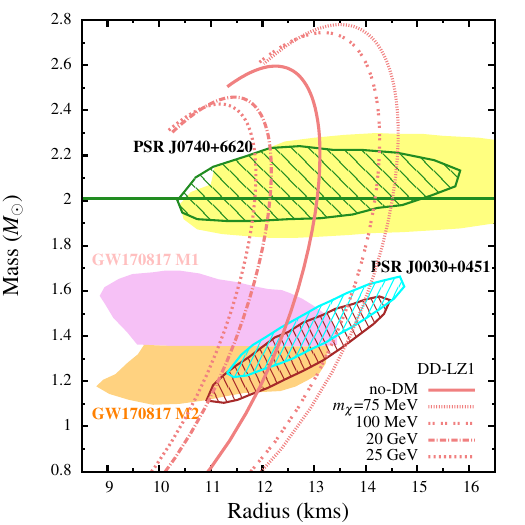}\protect\label{mr_mchiDDLZ1}}
\caption{\it Variation of mass with radius of dark matter admixed neutron stars for different values of $m_{\chi}$ and maximum $k_F^{\chi}$ with hadronic models (a) PK-DD and (b) DD-LZ1.}
\end{figure}

\begin{figure}[!ht]
\centering
\subfloat[PK-DD]{\includegraphics[width=0.5\textwidth]{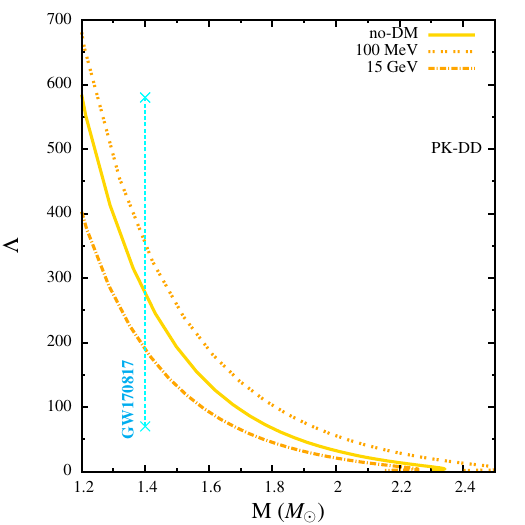}\protect\label{LamM_mchiPKDD}}
\subfloat[DD-LZ1]{\includegraphics[width=0.5\textwidth]{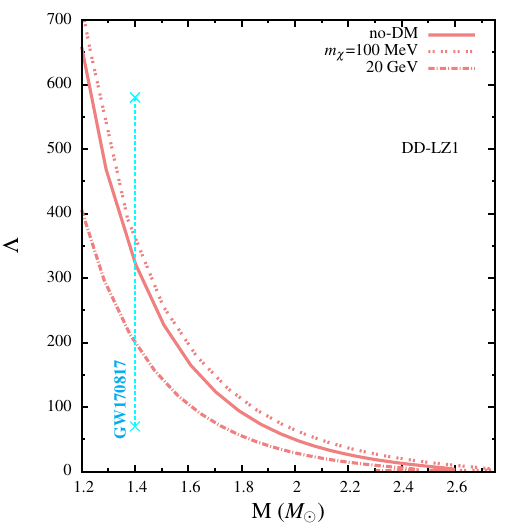}\protect\label{LamM_mchiDDLZ1}}
\caption{\it Variation of tidal deformability with mass of dark matter admixed neutron stars for different values of $m_{\chi}$ and maximum $k_F^{\chi}$ with hadronic models (a) PK-DD and (b) DD-LZ1.}
\end{figure}

In Figs.~\ref{mr_mchiTM1}, \ref{mr_mchiGM1}, \ref{mr_mchiNL3}, \ref{mr_mchiPK1}, \ref{mr_mchiDDMEX}, \ref{mr_mchiDD2}, \ref{mr_mchiTW99}, \ref{mr_mchiDDME2}, \ref{mr_mchiPKDD} and \ref{mr_mchiDDLZ1} we show the maximum and minimum values of $m_{\chi}$ for which the DMANS configurations with maximum $k_F^{\chi}$ ($\rho_{\chi}$) can satisfy all the astrophysical constraints on the mass-radius variation for hadronic models TM1, GM1, NL3, PK1, DD-MEX, DD2, TW99, DD-ME2, PK-DD and DD-LZ1, respectively. For better understanding we also show the results for two more values of $m_{\chi}$ - one bellow the minimum and one above the maximum limits for each model in order to obtain a moderately clear range of $m_{\chi}$. It is seen that for a value of $m_{\chi}$ below the minimum limit, the result is inconsistent with the GW170817 data while the choice of $m_{\chi}$ above the maximum, leads to the violation of the NICER data for PSR J0030+0451. The obtained allowed range of $m_{\chi}$ for the DMANSs is then tested in the $M-\Lambda$ plane with respect to the constraint on the tidal deformability of 1.4 $M_{\odot}$ neutron star ($\Lambda_{1.4}$) obtained from the GW170817 data in Figs.~\ref{LamM_mchiTM1}, \ref{LamM_mchiGM1}, \ref{LamM_mchiNL3}, \ref{LamM_mchiPK1}, \ref{LamM_mchiDDMEX}, \ref{LamM_mchiDD2}, \ref{LamM_mchiTW99}, \ref{LamM_mchiDDME2}, \ref{LamM_mchiPKDD} and \ref{LamM_mchiDDLZ1} for the hadronic models TM1, GM1, NL3, PK1, DD-MEX, DD2, TW99, DD-ME2, PK-DD and DD-LZ1, respectively. Except for NL3, the obtained allowed range of $m_{\chi}$,  in terms of the different astrophysical constraints, is same in both the $M-R$ and $M-\Lambda$ planes. For the NL3 model the lower limit of $m_{\chi}$=500 MeV satisfy all the astrophysical constraints in the $M-R$ plain as seen from Fig.~\ref{mr_mchiNL3} but Fig.~\ref{LamM_mchiNL3} shows that with this value of $m_{\chi}$=500 MeV the result offshoots the upper bound on $\Lambda_{1.4}$ obtained from GW170817 data and combining the joint results of Figs.~\ref{mr_mchiNL3} and \ref{LamM_mchiNL3}, we find that the minimum value of 
$m_{\chi}$ for maximum $k_F^{\chi}$=0.07 GeV is 1 GeV in order to satisfy all the astrophysical constraints.

\begin{figure}[!ht]
\centering
\subfloat[$M_{max}$ vs $m_{\chi}$]{\includegraphics[width=0.33\textwidth]{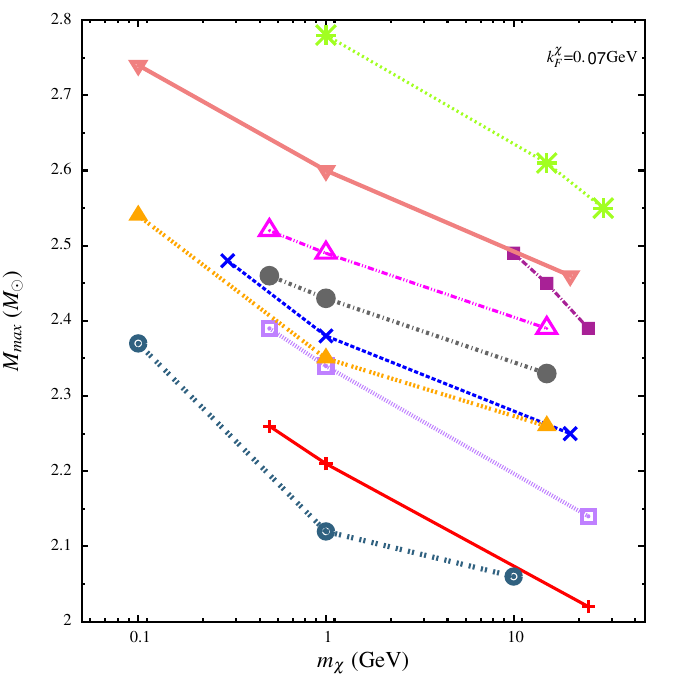}\label{mmchi_0p06}}
\hfill
\subfloat[$R_{1.4}$ vs $m_{\chi}$]{\includegraphics[width=0.33\textwidth]{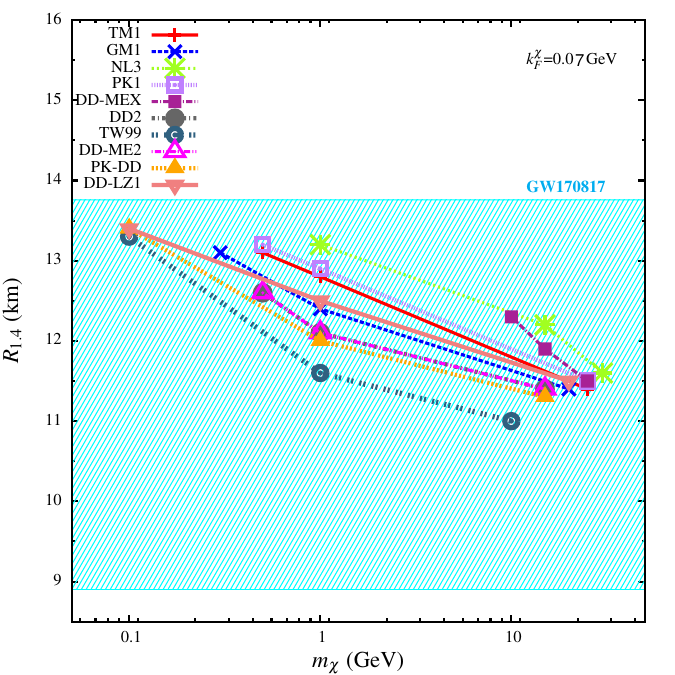}\label{rmchi_0p06}}
\hfill
\subfloat[$\Lambda_{1.4}$ vs $m_{\chi}$]{\includegraphics[width=0.33\textwidth]{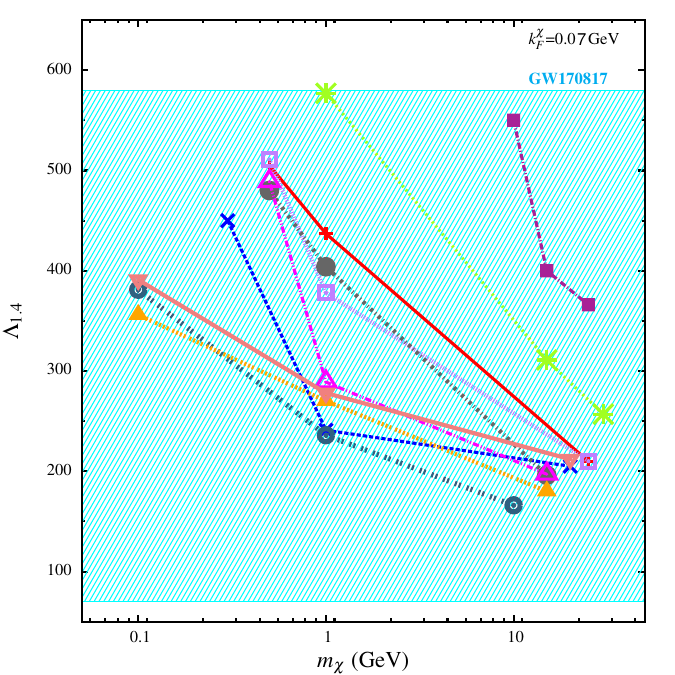}\label{Lmchi_0p06}}
\caption{\it Variation of (a) $M_{max}$, (b) $R_{1.4}$ and (c) $\Lambda_{1.4}$ with $m_{\chi}$ of dark matter admixed neutron stars at maximum $k_F^{\chi}$ for different hadronic models within the range of fulfillment of all the astrophysical constraints.}
\end{figure}

 For further understanding of the allowed range of $m_{\chi}$ that yields reasonable DMANSs that can successfully satisfy the constrained properties like maximum mass $M_{max}$, $R_{1.4}$ and $\Lambda_{1.4}$, we depict the individual variation of these quantities with $m_{\chi}$ in Figs. \ref{mmchi_0p06}, \ref{rmchi_0p06} and \ref{Lmchi_0p06}, respectively. For convenience we also compare the respective constraints in the same figure. In Figs.~\ref{mmchi_0p06}, \ref{rmchi_0p06} and \ref{Lmchi_0p06} we show with each model, the results for the two extreme values of $m_{\chi}$ that signifies the allowed range of $m_{\chi}$ for which the DMANS satisfy all the astrophysical constraints. Consistent with our previous works \cite{Sen:2021wev, Guha:2021njn} we find that for any model, lighter fermionic DM results in more massive DMANSs with larger radius. In Table \ref{tab:max} we present the allowed range of $m_{\chi}$ thus obtained for the maximum DM fraction with the ten chosen models.

\begin{table}[!ht]
\caption{The range of $m_{\chi}$ for which the dark matter admixed neutron stars at maximum $k_F^{\chi}$ satisfy all the astrophysical constraints on the structural properties of compact stars.}
\setlength{\tabcolsep}{100.0pt}
\begin{tabular}{cccccc}
\hline
\hline
Model & $m_{\chi}$ (GeV) \\
\hline
TM1     & 0.5 $-$ 25 \\
GM1     & 0.3 $-$ 20 \\
NL3     & 1.0 $-$ 30   \\
PK1     & 0.5 $-$ 25  \\
DD-MEX  & 1.0 $-$ 25   \\
DD2     & 0.5 $-$ 15  \\
TW99    & 0.1 $-$ 10  \\
DD-ME2  & 0.5 $-$ 15  \\
PK-DD   & 0.1 $-$ 15  \\
DD-LZ1  & 0.1 $-$ 20  \\
\hline
\hline
\end{tabular}
\label{tab:max}
\end{table}

\subsection{Dark matter admixed neutron stars with minimum $k_F^{\chi}$=0.01 GeV}

We now proceed to obtain our results with the minimum value of $k_F^{\chi}$=0.01 GeV which corresponds to the minimum DM density $\rho_{\chi}^{min}$(= 4.4 $\times$ 10$^{-6}$ fm$^{-3}$). In the same way as in case of the maximum $k_F^{\chi}$, we try to obtain the allowed range of $m_{\chi}$ required to obtain reasonable DMANSs configurations for the minimum value of $k_F^{\chi}$.

\begin{figure}[!ht]
\centering
\subfloat[$M_{max}$ vs $m_{\chi}$]{\includegraphics[width=0.333\textwidth]{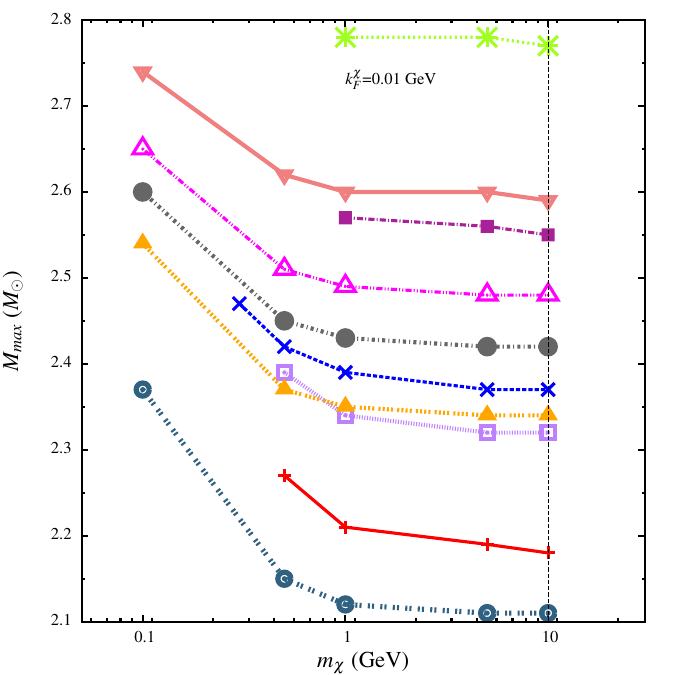}\label{mmchi_0p01}}
\hfill
\subfloat[$R_{1.4}$ vs $m_{\chi}$]{\includegraphics[width=0.333\textwidth]{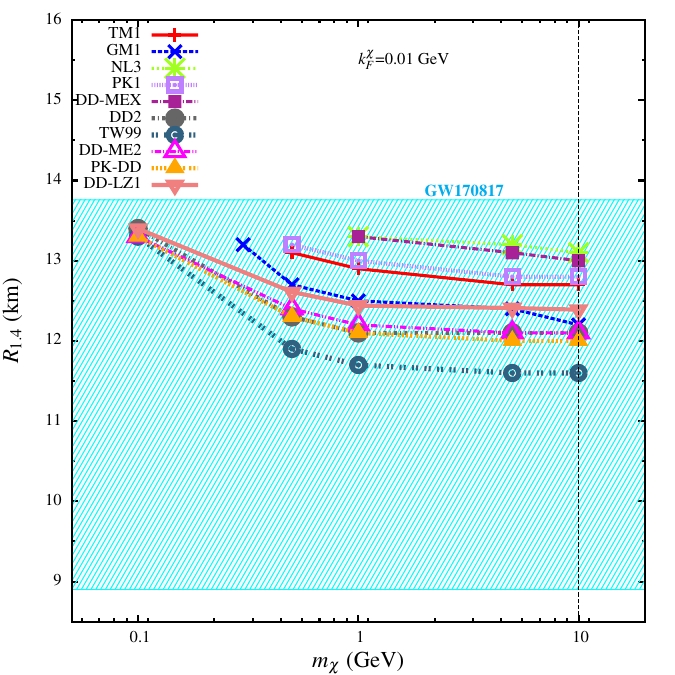}\label{rmchi_0p01}}
\hfill
\subfloat[$\Lambda_{1.4}$ vs $m_{\chi}$]{\includegraphics[width=0.333\textwidth]{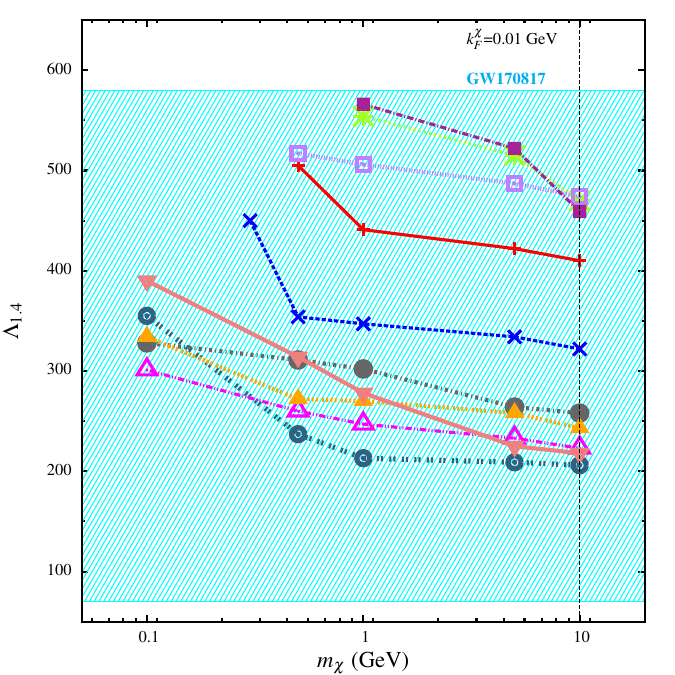}\label{Lmchi_0p01}}
\caption{\it Variation of (a) $M_{max}$, (b) $R_{1.4}$ and (c) $\Lambda_{1.4}$ with $m_{\chi}$ of dark matter admixed neutron stars at minimum $k_F^{\chi}$ for different hadronic models within the range of fulfillment of all the astrophysical constraints. The black dashed vertical line indicate saturation of the values of maximum mass, $R_{1.4}$ and $\Lambda_{1.4}$ at 10 GeV for all the hadronic models.}
\end{figure}

Similar to Figs.~\ref{mmchi_0p06}, \ref{rmchi_0p06} and \ref{Lmchi_0p06} obtained for maximum $k_F^{\chi}$, we present for the minimum $k_F^{\chi}$=0.01 GeV, in Figs.~\ref{mmchi_0p01}, \ref{rmchi_0p01} and \ref{Lmchi_0p01} the dependence of $M_{max}$, $R_{1.4}$ and $\Lambda_{1.4}$ on $m_{\chi}$ with respect to the constraints on these quantities. Interestingly, in the case of very low DM population, we find that the values of $M_{max}$, $R_{1.4}$ and $\Lambda_{1.4}$ saturate at a maximum value of $m_{\chi}$=10 GeV i.e, above this value of $m_{\chi}$ the structural properties of the DMANSs do not change for any of the hadronic models considered in the present work. Therefore in this case of minimum $k_F^{\chi}$ we do not obtain any particular upper bound on $m_{\chi}$ but a saturation value $m_{\chi}^{sat}$=10 GeV irrespective of the hadronic model considered to obtain the DMANS configurations. This is because with the lower DM population, the scenario is close to the no-DM case and under such circumstances the low DM content cannot bring any perceptible change to the structural properties of the star. For example the maximum mass of both the DMANS for $m_{\chi}$=10 GeV and the neutron star in the no-DM scenario is 2.32 $M_{\odot}$ for the PK1 model while it is 2.42 for the DD2 model. For lower $k_F^{\chi}$, $m_{\chi}$ saturates at a lower value compared to that for a higher value of $k_F^{\chi}$. So for $k_F^{\chi}$=0.07 GeV, the value of $m_{\chi}^{sat}$ is quite higher and beyond the maximum value of $m_{\chi}$ required to satisfy all the astrophysical constraints. Therefore in Table \ref{tab:min} we tabulate the minimum values of $m_{\chi}$ for which the DMANS at minimum $k_F^{\chi}$ satisfy all the astrophysical constraints.

\begin{table}[!ht]
\caption{The minimum value of $m_{\chi}$ for which the dark matter admixed neutron stars at minimum $k_F^{\chi}$ satisfy all the astrophysical constraints on the structural properties of compact stars.}
\setlength{\tabcolsep}{100.0pt}
\begin{tabular}{cccccc}
\hline
\hline
Model & $m_{\chi}^{min}$ (GeV)\\
\hline
TM1     & 0.5 \\
GM1     & 0.3 \\
NL3     & 1.0   \\
PK1     & 0.5  \\
DD-MEX  & 1.0  \\
DD2     & 0.1 \\
TW99    & 0.1 \\
DD-ME2  & 0.1 \\
PK-DD   & 0.1  \\
DD-LZ1  & 0.1  \\
\hline
\hline
\end{tabular}
\label{tab:min}
\end{table}

\begin{figure}[!ht]
\centering
{\includegraphics[width=0.8\textwidth]{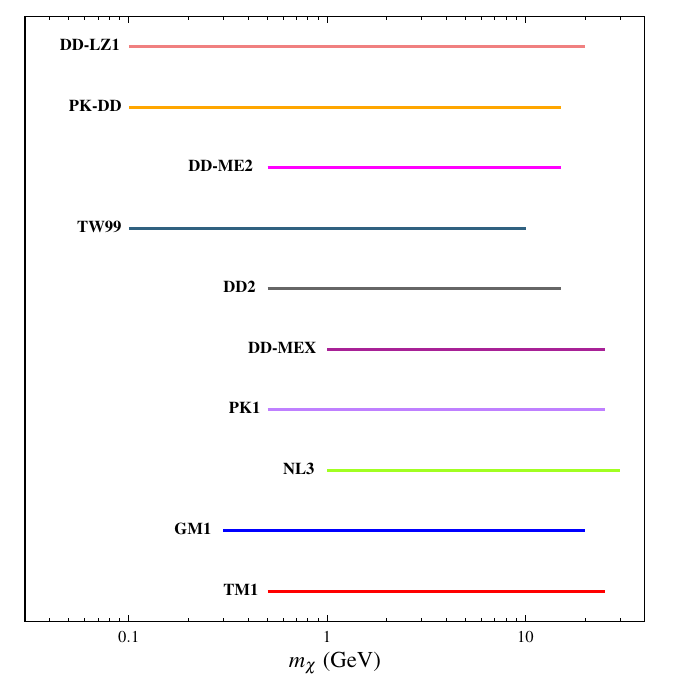}
\caption{\it The combined range of $m_{\chi}$ for which the dark matter admixed neutron stars satisfy all the astrophysical constraints on the structural properties of compact stars for all the hadronic models.}
\protect\label{Range_mmchi}}
\end{figure}

Thus combining the results of the Tables \ref{tab:max} and \ref{tab:min} we obtain a range of $m_{\chi}$ for which the DMANS satisfy all the astrophysical constraints within a wide range of $k_F^{\chi}$=(0.01 $-$ 0.07) GeV or wide range of DM fraction in neutron stars. We present this combined range of $m_{\chi}$ in Fig.~\ref{Range_mmchi}. It can be seen from Fig.~\ref{Range_mmchi} that for the models like NL3, TM1, PK1, and DD-MEX that do not or barely satisfy the constraints on $R_{1.4}$ and $\Lambda_{1.4}$ from GW170817 in the absence of DM (Fig.~\ref{mLamM_noDM}), comparatively massive DM is required to obtain reasonable (with respect to the various astrophysical constraints) DMANSs configurations. We also find that considering all the ten RMF hadronic models chosen for the present work, the combined range of $m_{\chi}\approx$ (0.1 $-$ 30) GeV for a wide range of $k_F^{\chi}$=(0.01 $-$ 0.07) GeV.

\section{Summary and Conclusion}
\label{Conclusion}

In the present work we aim to study the effects of feeble interaction between hadronic matter and fermionic DM via new physics scalar and vector mediators on the structural properties of the DMANSs in the light of the different astrophysical constraints. For the purpose we consider ten well-known RMF models to describe the pure hadronic matter. $m_{\chi}$, $m_{\phi}$ and $m_{\xi}$ are consistent with the self-interaction constraint from Bullet cluster while $y_{\phi}$ and $y_{\xi}$ are constrained by the present day relic abundance. We assume that both $\phi$ and $\xi$ contribute equally to the relic abundance and compute the equation of state and the structural properties of the DMANSs. In order to satisfy the various recent constraints like those from the massive pulsar PSR J0348+0432, the gravitational wave (GW170817) data and the results of NICER experiments for PSR J0030+0451 and PSR J0740+6620, we find that within the framework of the present work, the DMANSs may contain fermionic DM of mass in the range of $m_{\chi}\approx$ (0.1 $-$ 30) GeV corresponding to a wide range of fixed $k_F^{\chi}$=(0.01 $-$ 0.07) GeV. For the above mentioned mass range of DM, the DMANSs well satisfy the astrophysical constraints on structural properties of the compact stars. This range of $m_{\chi}$ can be considered to be potentially favorable in order to explain the possible existence of the DMANSs.

 For the hadronic models that yield larger radii corresponding to the low mass neutron stars in the no-DM scenario, interaction with comparatively heavier DM fermion is necessary in order to ensure that the DMANSs obtained with such models satisfy the radius constraints from both GW170817 and NICER data for PSR J0030+0451.


\section*{Acknowledgements}

Work of A.G. is supported by the National Research Foundation of Korea (NRF-2019R1C1C1005073). Work of DS is supported by the NRF research Grants (No. 2018R1A5A1025563).


\bibliography{ref}

\end{document}